\documentclass[journal]{IEEEtran}

\usepackage[superscript,biblabel]{cite}

      \usepackage{amssymb}
      \usepackage{amsmath}
      
      \usepackage[noend]{algpseudocode}
      
      \usepackage{easybmat}
      
      \usepackage{epsfig,graphicx,multirow,pdfpages}
      
      \usepackage{adjustbox}
      
      \usepackage[T1]{fontenc}
      \usepackage{textcomp}
      
      \usepackage[boxed,noend]{algorithm2e}
      
      \SetAlgoCaptionSeparator{.}
      \SetAlCapSkip{3pt}
      
      \usepackage{hyperref}
      \usepackage[nameinlink,noabbrev]{cleveref}
      
      \usepackage{enumitem}
      \usepackage{xcolor}
      
      \usepackage{amsthm}
      \theoremstyle{definition}
      \newtheorem{definition}{Definition}[section]

   \makeatletter 
   \newcommand\mynobreakpar{\par\nobreak\@afterheading} 
   \makeatother
   
      \makeatletter
      \def\@setcopyright{}
      \def\serieslogo@{}
      \makeatother

\makeatletter
\def\BState{\State\hskip-\ALG@thistlm}
\makeatother

\makeatletter

\def\mydashbox#1#2{%
\setbox0\hbox{#2}%
\dimen0\ht0
\advance\dimen0\dp0
\setbox2\vbox to \dimen0{{\color{#1}\leaders\vbox{\vskip2pt\hrule height 2pt width .3pt}\vfill}}%
\ht2=\ht0
\dp2=\dp0
\box2
\unhbox0
}

   \newcommand{\etal}{\emph{et al. }}
   \newcommand{\dash}{\,--\,}
   \newcommand{\doi}[1]{doi:\href{http://dx.doi.org/#1}{\texttt{#1}}}

\usepackage[font=footnotesize]{caption}
\usepackage{float}
\newcommand{\myfigure}[4]{
\begin{figure*}[#2]
\captionsetup{width=0.85\textwidth, justification=raggedright}
\setlength{\abovecaptionskip}{3pt}
\centering
#4 
\caption{#3} 
\label{fig#1} 
\end{figure*}
}

\newcommand{\myfigureexpl}[5]{
\begin{figure*}[#2]
\captionsetup{width=0.85\textwidth, justification=raggedright}
\setlength{\abovecaptionskip}{3pt}
\centering
#5 
\caption{#3} 
\setlength{\abovecaptionskip}{1pt}
\caption*{#4} 
\label{fig#1} 
\end{figure*}
}

\def\NTPname{Note to Practitioners}

\def\NtP{\normalfont
    \if@twocolumn
      \@IEEEabskeysecsize\bfseries\textit{\NTPname}---\relax
    \else
      \bgroup\par\addvspace{0.5\baselineskip}\centering\vspace{-1.78ex}\@IEEEabskeysecsize\textbf{\NTPname}\par\addvspace{0.5\baselineskip}\egroup\quotation\@IEEEabskeysecsize
    \fi\@IEEEgobbleleadPARNLSP}

\def\endNtP{\relax\ifCLASSOPTIONconference\vspace{0ex}\else\vspace{1.34ex}\fi\par\if@twocolumn\else\endquotation\fi
    \normalfont\normalsize}

        \renewcommand{\algocf@Vsline}[1]{%
                \strut\par\nointerlineskip%
                \algocf@push{\skiprule}%
                \hbox{%
            \mydashbox{black}{%
                        \vtop{\algocf@push{\skiptext}%
                        \vtop{\algocf@addskiptotal\advance\hsize by -\skiplength #1}}%
                }%
            }
                \algocf@pop{\skiprule}%
        }
\makeatother

\begin{document}

\title{Freeform Assembly Planning}

\author{Matthew~K.~Gelber,
        Greg~Hurst,
        and~Rohit~Bhargava
\thanks{This work was supported in part by the Beckman Institute for Advanced Science and Technology via its seed grant program.}
\thanks{M. Gelber is with the Beckman Institute for Advanced Science and Technology, and the Department of Bioengineering, University of Illinois at Urbana-Champaign, Urbana, IL, 61801, \href{mailto:mgelber2@gmail.com}{\texttt {mgelber2@gmail.com}}.}
\thanks{G. Hurst is with Wolfram Research in Champaign, IL, 61820, \href{mailto:ghurst@wolfram.com}{\texttt {ghurst@wolfram.com}}.}
\thanks{R. Bhargava is with the Beckman Institute for Advanced Science and Technology and the Departments of Chemistry, Chemical and Biomolecular Engineering, Mechanical Science and Engineering, and Electrical and Computer Engineering, University of Illinois at Urbana-Champaign, Urbana, IL.}
\thanks{Correspondence to: Rohit Bhargava, Beckman Institute for Advanced Science and Technology, 405 N. Mathews Ave, Urbana, IL 61801; Telephone: 217-265-6596; Fax: 217-265-0246; Email: \href{mailto:rxb@illinois.edu}{\texttt {rxb@illinois.edu}}.}
}


\maketitle

\begin{abstract}
\boldmath
3D printing enables the fabrication of complex architectures at multiple length scales by automating large sequences of additive steps. The increasing sophistication of printers, materials, and generative design promises to make geometric complexity a non-issue in manufacturing; however, this complexity can only be realized if a design can be translated into a physically executable sequence of printing operations. We investigate this planning problem for freeform direct-write assembly, in which filaments of material are deposited through a nozzle translating along a 3D path to create sparse, frame-like structures. We enumerate the process constraints for different variants of the freeform assembly process and show that, in the case where material stiffens \emph{via} a glass transition, determining whether a feasible sequence exists is NP-complete. Nonetheless, for topologies typically encountered in real-world applications, finding a feasible or even optimal sequence is a tractable problem. We develop a sequencing algorithm that maximizes the fidelity of the printed part and minimizes the probability of print failure by modeling the assembly as a linear, elastic frame. We implement the algorithm and validate our approach experimentally, printing objects composed of thousands of large-aspect-ratio sugar alcohol filaments with diameters of 100-200 $\mu$m.
\end{abstract}

\begin{NtP}
Extrusion-style 3D printers typically pattern material in a series of 2D layers, but they can be also be programmed to deposit material along 3D paths in a ``freeform'' fashion. However, programming a printer to operate in this way requires consideration of constraints related to collision and stability. For large designs, finding an optimal or even feasible plan with respect to these constraints requires automated planning. We address a hard version of this problem in which any joint in the frame can only support one cantilever at a time. We develop an exact algorithm that maximizes the robustness of the printing plan, and validate it by printing complex freeform designs. The assembly planner allows the freeform process to be applied to arbitrarily complex parts, with applications ranging from tissue engineering and microfluidics at the micrometer scale, to vascularized functional materials and soft robots at the millimeter scale, to structural components at the meter scale. This approach removes a major bottleneck in the workflow for freeform assembly, allowing scientists and engineers to automatically translate complex freeform designs into optimal printing plans.
\end{NtP}

\begin{IEEEkeywords}
assembly planning, additive manufacturing, 3D printing, freeform 3D printing, wireframe printing, freeform assembly, direct-write assembly, manufacturing constraints, stability constraints, collision constraints
\end{IEEEkeywords}

\IEEEpeerreviewmaketitle

\section{Introduction}
\label{intro}

\IEEEPARstart{I}{n} 3D printing, complexity is sometimes said to be ``free'', because every assembly step performed by the printer is automated, and the process permits features such as internal voids that cannot be created by subtractive approaches. Though some types of complexity are certainly more accessible using additive manufacturing, complexity is not free; failures and artifacts still occur, and some geometries are physically unconstructable on some printers. In some cases, these unconstructable designs can be made constructable by increasing the degrees of freedom available to the printer. For example, a physical barrier to 3D printing a given design might be overcome by allowing material to be deposited along 3D paths instead of layer-by-layer, or by allowing the part to be rotated mid-print to change the direction of the gravitational force on it. Though additional degrees of freedom in the process increase the accessible design space, they also increase the number of ways in which a part could potentially be printed as well as the number of ways in which a print could fail. In some cases, there may exist a sequence of operations that would successfully produce a given design, but there is a computational barrier to determining that sequence. In this work, we explore a variant of 3D printing where this is the case and demonstrate an approach for overcoming the computational barrier by means of an efficient state-space graph search.
   
In most forms of 3D printing, material is deposited, polymerized, or sintered in a series of layers.\cite{1} This approach is robust and straightforward to implement. For some types of parts, however, greater speed and fidelity can be achieved by depositing material along paths that are not constrained to a supporting surface. This is done by extruding material from a nozzle moving in three dimensions, with a mechanism to prevent the material from deforming after extrusion. In some implementations, the extruded material is supported by a shear-thinning fluid\cite{2,3} or $\mu$m-scale gel particles.\cite{4,5,6} If the extruded material can be made to rapidly stiffen after extrusion, a supporting reservoir may be unnecessary, and the material can be deposited in free space. Stiffening mechanisms include UV-induced polymerization\cite{7,8}, thermally induced polymerization\cite{9}, solvent evaporation\cite{10}, shear-thinning\cite{11}, laser sintering\cite{12}, and, in the case of thermoplastics\cite{13,14,15,16,17}, metals\cite{18}, and glasses\cite{19,20,21}, cooling below the melting or glass transition temperature. Freeform patterning, layer-by-layer patterning, and patterning directly onto a flat or curved surface have been broadly termed \emph{direct-write assembly} in the scientific literature. We use the term freeform assembly to refer to the subset of direct-write assembly in which the patterned material is not constrained to lie on a particular plane or surface. In more recent work this is sometimes called ``wireframe'' 3D printing; we use the term ``assembly'' in recognition of the earlier work on direct-write assembly as well as to emphasize the connection between this problem and the well-established field of assembly planning.\cite{3}
   
Freeform assemblies can be concisely represented using a graph data structure. We refer to the nodes of this graph as \emph{joints} and to the edges as \emph{beams}. Each joint is specified by an \emph{xyz} coordinate. Each beam is specified by a 3D path connecting its 2 joints and by a thickness, which may vary along this path. Choosing an order in which to print the beams can be formulated as an assembly planning problem. Here we consider assembly plans that are linear and monotone.\cite{22} In a linear assembly plan, exactly one part is attached to the rest of the assembly at each step. In a monotone assembly plan, once added, no part is ever removed from the assembly. These constraints are appropriate for freeform 3D printers that print using one nozzle at a time and cannot remove any material that has already been printed. A linear, monotone plan is also called a sequence.

The problem of generating sequences for ``wireframe'' printing has received some recent attention from the computer graphics community. Wu \etal used a ``peeling'' approach to find collision-free sequences for wireframe printing using a 5-axis printer.\cite{16} Huang \etal examined a planning optimization problem for a printer based on a 6-axis robotic arm. Their planner uses a heuristic to avoid collision and also optimizes against a cost function related to the robustness and speed of the print.\cite{17} However, we find that the constraints and optimality criteria in these processes are not appropriate for ours. We develop an algorithm for our problem starting from a classical assembly planning framework and discuss ways in which this older body of work can simplify approaches to the problems formulated by Wu and Huang.

The essence of most assembly planning approaches is a search through the assembly state graph.\cite{23} In this graph, each state is either a single part or a set of parts that have been placed in their final configuration with respect to each other. The graph is initialized with states corresponding to single parts, and new states are generated from the union of existing states. An example assembly graph for a freeform assembly is shown in \cref{fig2}. For assemblies containing a sufficiently large number of parts, generating the entire assembly state graph is computationally prohibitive. It is only necessary to search the assembly graph for a path from the completely disassembled (empty) state to the completely assembled (full) state.\cite{24} The search may be initiated from either of these states. If starting from the empty state, a part added at one step may obstruct the placement of another part at a later step. Thus a forward search can result in dead-end states from which the assembly cannot be finished. For this reason, the graph search is typically begun from the full state. New states are generated by separating pairs of subassemblies, generating a disassembly sequence. Reversing the disassembly sequence gives the forward assembly sequence. Separating a subassembly will never create a new obstruction for the remaining parts; thus, if collision and connection are the only constraints, this approach will never create a dead-end state. Under stability constraints, however, it is possible to generate dead-end states even in the disassembly graph search.

The problem of assembly sequencing under stability constraints has been investigated for block towers\cite{25,26}, frame structures\cite{27,28}, and mechanical assemblies for which stability is a function of orientation.\cite{29,30} In these planning approaches, the problem of dead ends is avoided by allowing scaffolding or reorientation. Unstable states are penalized with a scaffolding or reorientation cost, but the planner does not backtrack in search of a sequence that minimizes this cost. The work of Huang \etal\cite{17} is an exception: the planner performs a stability test at each state and backtracks if the state fails the test. This required an additional divide-and-conquer approach to achieve an acceptable run time. Ideally, a planner would be able to find, without backtracking, sequences that contain no unfinishable or unconstructable states. We prove that, for the stability criterion appropriate for materials that stiffens \emph{via} a reversible glass transition, the decision version of this problem is NP-complete. However, for the types of geometries one would commonly encounter, we show that sequences that are feasible and even optimal under certain cost functions can be found efficiently. Lastly, we validate our approach by physically printing large assemblies composed of thousands of beams.

   \section{Problem Definition}

   \subsection{Process Constraints}
   \label{constraints}
   
   Different freeform assembly processes are subject to different subsets of the following constraints. If beam $A$ must be printed before beam $B$, we say that $A$ precedes $B$ ($A \prec B$), or that $B$ succeeds $A$ ($B \succ A$).
   
  {\bf Directionality:} No beam can be printed in a way that would cause the nozzle to collide with that beam during printing. For the 3-axis printer considered here, vertical beams must be printed bottom to top. Some beams are unprintable from either direction, and any assembly containing such a beam are unconstructable. If a beam is directed such that it must be printed starting from joint $p$, it succeeds at least one other beam incident to joint $p$.
   
   {\bf Collision:} No beam can be printed that would cause the nozzle to collide with a previously printed beam. If the volume swept out by the nozzle while printing beam $A$ intersects any part of beam $B$, then ($A \prec B$).
   
   {\bf Connection:} Each successive beam must start from a joint on the substrate or a joint to which a previously printed beam is incident. Any beam containing non-grounded joints $p$ and $q$ must succeed at least one of the other beams incident to joint $p$ or joint $q$.
   
   \myfigureexpl{1}{t}{Freeform assembly constraints}{(isometric and top-down views)}{
     \includegraphics[scale = .28]{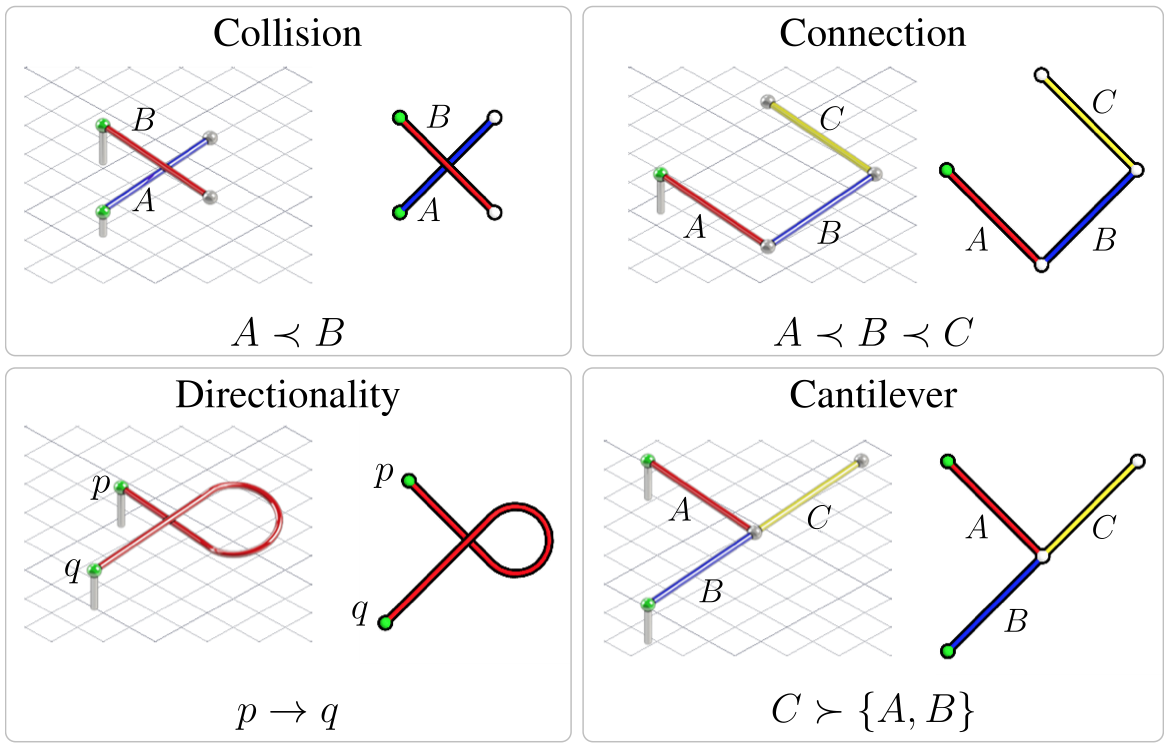}
   }
   
   {\bf Cantilever:}
   When stiffening is achieved by freezing or cooling below the glass transition temperature, special consideration must be given to the effect of heating an existing joint. When fusing a new beam to an existing joint, the heat of the nozzle may cause the joint to become fluid and unable to sustain a bending moment (Supplementary video 1). Any cantilevered beams attached to the joint will pivot due to the moment caused by their own weight. This introduces a cantilever constraint: new beams cannot be connected to joint $p$ if any previously printed beam incident to joint $p$ is cantilevered. It may be possible to limit heat flow to the joint and prevent pivoting -- for example, by depositing extra material at the joint, or by offsetting the nozzle slightly -- but we will treat this as an absolute constraint in order to obtain a highly robust assembly sequence. Any beam cantilevered about joint $p$ must succeed all other beams incident to $p$. Note that, unlike the other constraints, this constraint only applies to some subset of the possible assembly states.
   
All freeform assembly processes are subject to directionality and collision constraints. Free-standing assemblies are subject to the connection constraint, and assemblies in which stiffening is achieved by a reversible glass or phase transition are subject to the cantilever constraint. In the absence of the cantilever constraint, finding a feasible sequence is analogous to task planning under AND/OR constraints, and can be done in linear time with respect to the number of beams\cite{31}. Here we show that the addition of the cantilever constraint renders the decision problem NP-complete.
   
As in classical assembly planning, our approach will use a state-space search. Though assembly sequencing is often done by finding a disassembly sequence and reversing it, we will use a forward search.  For a process with only 3 degrees of freedom, the amount of collision computation saved by performing a disassembly search compared to an assembly search is not significant. Also, while a disassembly planner must finish the entire sequence before the physical printing process can begin, a forward planner can be run concurrently with a printer, generating the sequence on the fly. If the printing time is $t_1$, and the planning time is $t_2$, this can reduce the total time from $t_1 + t_2$ to as little as $\min(t_1, t_2)$, i.e., up to a factor of 2. All of the arguments made apply to the disassembly problem as well, and an essentially identical approach could be applied to generate a disassembly sequence.

   \subsection{Computational Complexity}
   \label{complexity}
   
   We seek a sequence of beams and a printing direction for each beam that are consistent with all of the process constraints. Collision, connection, and directionality constraints are the same regardless of the sequence of states, but the cantilever constraint is a function of the assembly state itself, and different cantilever constraints will be encountered for different sequences. This renders the problem much more difficult, because a na\"ive graph search may reach a dead end state whether the search is initiated from the full state or the empty state. Reaching a dead end requires backtracking to a previous state to resume the search. In the worst case, the search will enumerate the entire graph.
   
   \myfigureexpl{2}{t}{Freeform Assembly State Graph}{Since the assembly is monotonic, each new state is the union of an existing state and a single beam. The single part is not shown in the graph, so each state above is equivalent to an OR node.}{
     \includegraphics[scale = .31]{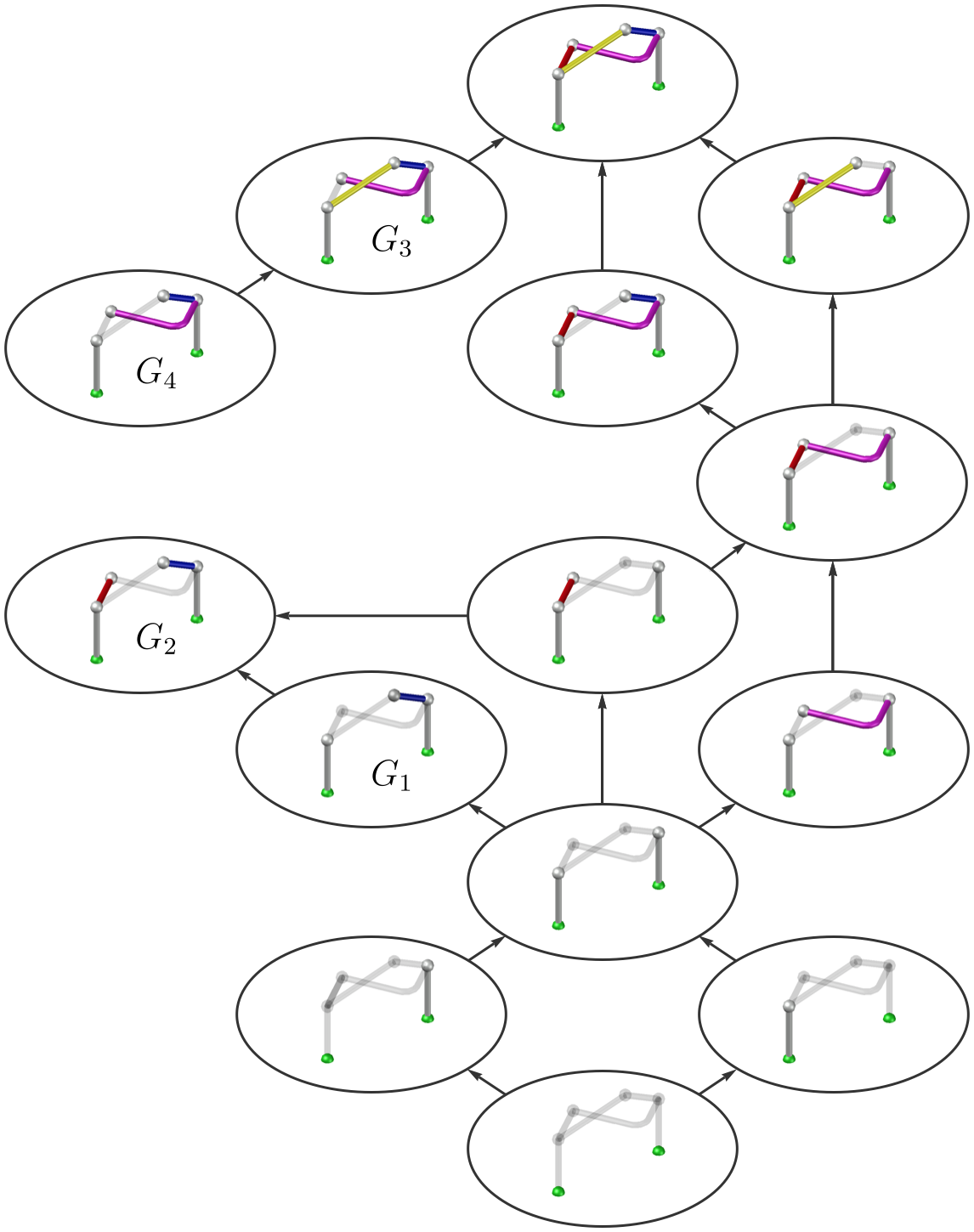}
   }

State $G_4$ in \cref{fig2} can be immediately identified as a disassembly dead end because there is a joint supporting 2 cantilevers, resulting in conflicting precedence constraints; the magenta and blue beams must precede each other due to the cantilever constraint, which makes it impossible to assemble both of them. Similarly, state $G_2$ can be immediately identified as an assembly dead end because assembling either the yellow or magenta beams would violate the cantilever constraint. However, there can exist states in which no conflict is immediately identifiable, but which are nonetheless dead ends. An assembly search may reach a state from which the full state is unconstructable, as occurs in state $G_1$, while a disassembly search may encounter a state that is unconstructable from the empty state, as occurs in state $G_3$. In order to search the graph efficiently, one must have a means of avoiding these branches from which the goal state is unreachable. Determination of a feasible sequence is polynomial time reducible to the decision problem of state reachability, $Reachable(G,H)$. This routine returns true if there exists a sequence of beams that can be successively added to assembly state $G$ to create assembly state $H$ without violating any process constraints. The reduction can be done as described in \cref{sequence}.

   \begin{algorithm}
     \SetArgSty{}
     \DontPrintSemicolon
     \SetAlCapFnt{\footnotesize \sf}
     \SetAlCapNameFnt{\footnotesize \sf}
     \caption{A polynomial time reduction from enumeration of a feasible sequence (search problem) to $Reachable(G,H)$ (decision problem). The sequence is recorded in $seq$}
     \label{sequence}
     $G \leftarrow \text{empty assembly state}$\;
     $H \leftarrow \text{full assembly state}$\;
     $B \leftarrow \text{beams}$\;
     $seq \leftarrow \{\}$\;
     \While{$|B| > 0$}{
       \For{each $b$ in $B$}{
         \If{$Reachable(G \cup \{b\}, H)$}{
           append $b$ to $seq$\;
           delete $b$ from $B$\;
           $G \leftarrow G \cup \{b\}$\;
         }
       }
       \If{no element was removed from $B$}{
         \Return unconstructable
       }
     }
     \Return $seq$\;
   \end{algorithm}
   
   We now prove that state reachability is NP-complete by polynomial time reduction from circuit satisfiability (circuit-SAT). The circuit-SAT problem asks if, given an arbitrary Boolean expression, there exists a satisfying assignment to each variable such that the expression evaluates to true. We will show that, for any instance of circuit-SAT, it is possible to design an assembly such that, if the full assembly state is reachable from the empty state, then the corresponding instance of circuit-SAT is satisfiable. The logic is as follows: we design an assembly that can be fully assembled if and only if one particular circuit ``output'' beam can be assembled. If the circuit output beam and thus the full assembly state can be assembled, then $Reachable(Empty State, Full State)$ will return true. If the circuit output beam cannot be assembled, then the full assembly state is unreachable as well, and $Reachable(Empty State, Full State)$ will return false. Each variable in the instance of circuit-SAT represented by this assembly will correspond to a specific beam. If a variable is true, then its corresponding beam must be assembled before the output beam. If a variable is false, then its corresponding beam must be assembled after the circuit output beam. If the circuit output beam can be assembled, its value is true, and there exists a satisfying assignment for the circuit that the assembly instantiates.
   
This scheme transforms a set of Boolean constraints between variables into a set of precedence constraints between beams. In order to represent any instance of circuit-SAT, we need a method of imposing arbitrary precedence constraints between the variable beams. To do so we introduce additional beams to create ``gadgets'' and ``wires'', as in classical circuit-SAT proofs. Each wire corresponds to a single beam. If a wire carries a value of 1 or true, that beam is present, i.e., has been assembled in that state. If a wire carries a value of 0 or false, that beam is absent, i.e., has not been assembled in that state. Gadgets are composed of several beams that create precedence relationships between wires that connect them. These precedence relationships will constrain the permissible assembly states of each gadget in a way that mirrors the permissible states of logic gates.
   
Any Boolean operation or logic gate can be represented using either of 2 universal gadgets, NAND or NOR, and a splitter gadget. The splitter gadget creates two output wires with the same value as the input wire; this is necessary for this proof because wires are represented by single beams, which can only connect one gadget input to one gadget output. We will implement the NOR and splitter logic gates by introducing a set of beams such that, for each state of the gate, there is exactly one state of the gadget that would enable assembly of the output beam. Before describing the gadgets, we begin with some preliminary definitions.
   
   \begin{definition}
     A beam \emph{path} is a path in the sense of graph theory, i.e., a connected acyclic set of beams.
   \end{definition}
   \begin{definition}
     A joint is \emph{grounded} if it is on the substrate.
   \end{definition}
   \begin{definition}
     A joint $p$ is \emph{stable} there are 2 or more paths from $p$ to a grounded joint that intersect only at $p$. A path is stable if all the joints in it are stable.
   \end{definition}
   \begin{definition}
     A \emph{connected} subassembly is a connected component in the sense of graph theory, where the beams are the graph edges and the joints are the nodes. I.e., for each connected subassembly, there is a path between any 2 joints in that connected subassembly.
   \end{definition}
   
The gadgets are shown in \cref{fig3} in a top-down view. If beam $A$ crosses over beam $B$, then $B \succ A$ in accordance with the collision constraint. Joints are represented by dots. Green joints are grounded. Wires are represented by white beams, and gadgets are represented by red, blue, and magenta beams. For both gadgets, there is a loop connected to a grounded joint. Due to the cantilever constraint, one half (the red or blue path) of the loop must be assembled and made stable before the other half is assembled. The only way to stabilize the red or blue path is to assemble the magenta path and connect it to the circuit output beam. Thus, exactly one path in the loop must be assembled before the circuit output beam, and the other path in the loop must be assembled after the circuit output beam. This XOR relationship can be exploited to yield both a splitter and a NOR gate by designing the gates such that inputs or outputs must precede or follow some portion of one of the red or blue paths. Note that this state constraint is only possible if the cantilever constraint applies; without it, only AND/OR precedence relations between states are possible, and the problem is in P.
   
   \myfigureexpl{3}{t}{Gadgets and assembly design for arbitrary boolean circuits}
   {NOR gadget - Exactly one of the red or blue paths must precede the output beam, which is shown in white crossing the magenta beam.\\
Splitter - As in the NOR gadget, exactly one of the red or blue paths must precede the output beam, which is shown in white crossing the magenta beam.\\
X NOR X - The magenta beams are stabilized by the assembly of a beam that is directed out of the end joint of the output beam. This ensures that the magenta beams cannot be stabilized by connection to each other. Hatches indicate that a beam is assembled; \\0 indicates true, and 1 indicates false.\\
Circuit Layout - Each box can represent either a splitter or a NOR gate, and so has at most 2 inputs and at most 2 outputs. If the gadgets are assembled top-to-bottom, starting at the leftmost column and proceeding to the right, then no magenta beam will ever prevent assembly of any wire beam.}{
     \includegraphics[scale = .3]{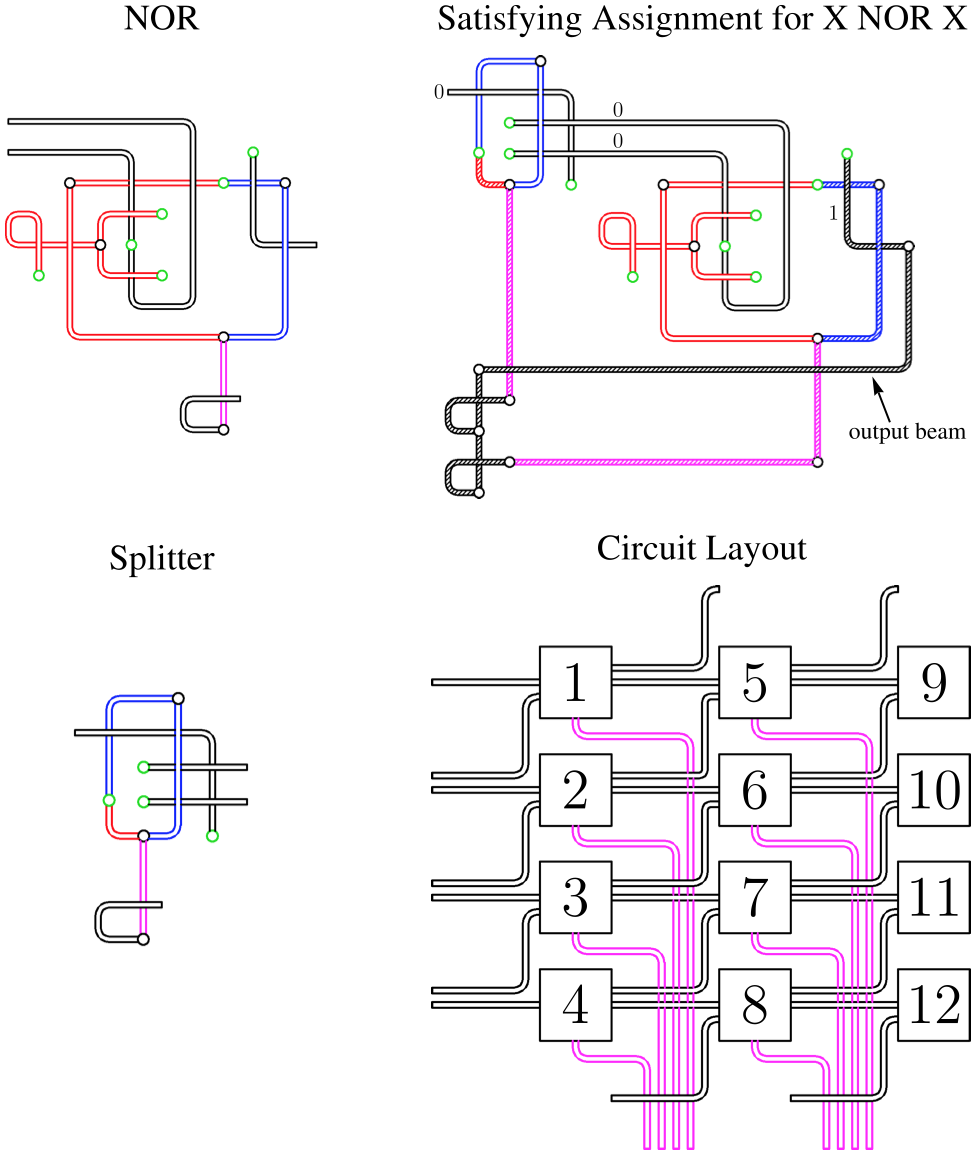}
   }
   
   In order to connect to the circuit output beam, the magenta gadget beams must cross over other wire beams, implying a precedence relation. Furthermore, the magenta beams must sometimes follow some of the input or output beams of the gadget from which they originate. However, this does not prevent instantiation of an arbitrary circuit. All magenta beams, but not all wire beams, must be assembled before the circuit output beam if the assembly is to be constructable. Therefore we must ensure that the magenta beams can be assembled before any wire beams that they cross, for any state of the circuit. Suppose the gadgets are arranged on a grid, and the magenta beams are routed to the right of their respective gadgets and then down to the bottom of the diagram, as shown in \cref{fig3}. The gadget states are assembled column-by-column, starting from the left, and each column is assembled starting from the top and proceeding down. In this way, the magenta beams can always be assembled before any wire beams they cross.
   
The circuit-SAT problem allows crossing wires, but in this proof, crossing beams imply a precedence relation. However, any circuit with crossovers can be transformed into an equivalent circuit without crossovers by using a crossover gate, which can be constructed using NOR gates.\cite{32} Thus although our circuit must have a planar embedding, a planar embedding can still any represent instance of circuit-SAT. If state reachability could be decided in polynomial time, then so could circuit-SAT. Therefore state reachability is NP-complete, as is deciding whether an assembly is constructable.
   
Although assembly constructability is NP-complete, we will develop a graph search algorithm that can quickly find assembly sequences that are not only feasible, but, for some criteria, optimal. Due to the sparse nature of freeform assembly designs, small deformations or positioning errors during the printing process can cause the print to fail catastrophically. When multiple beams are incident at a common joint, the actual position of the joint must be within a critical distance of the planned position. If the error between the planned and actual joint positions is too great, subsequent beams will fail to fuse to the joint, causing the print to fail. Similarly, if the actual shape of a beam deviates from its planned shape, the set of collision constraints computed from the original design may be wrong; subsequent beams may collide with this aberrant beam, whereas a perfectly printed beam would not cause collision. Though one might expect gravity to be the dominant source of dimensional error in freeform assembly, for our process, deformation is primarily caused by mechanical coupling between the extruded material and the rest of the assembly. If the diameter of the beam is smaller than the diameter of the nozzle, the nozzle will pull on the assembly in the direction of its motion. If the diameter of the beam is greater than the diameter of the nozzle, the extrusion pressure will push the assembly down, perpendicular to the nozzle orifice. Unless the diameter of a beam is perfectly matched to that of the nozzle, the assembly will store some elastic strain energy as the beam is printed. Once the beam is complete, the flow is shut off, removing the load, and the nozzle is held in place to fuse the beam to any existing joints. Because the material at the tip is molten, the beam can move so as to dissipate the internal strain energy in the assembly. However, if the strain energy is large, the beam end joint can recoil far from its planned position. If the error $\varepsilon_q$ between the planned position and equilibrium position is too great, the beam will detach from any other beams incident at its end joint and/or fail to fuse to subsequent beams incident at that joint. An example is shown in \cref{fig4}. If the horizontal beam is printed starting from the helical beam on the left, the downward force due to extrusion causes elastic strain energy to be stored in the highly compliant helix. When the hot nozzle is held at the end joint, the horizontal beam detaches from the vertical beam at the right, and the left side of the assembly springs back to its equilibrium conformation. In contrast, if the horizontal beam is printed starting from the relatively stiff vertical beam on the right, the assembly stores considerably less strain energy, and $\varepsilon_q$ is small enough that the horizontal beam and the helical beam remain fused.
   
   \myfigureexpl{4}{t}{Joint positioning errors due to beam compliance}{When the horizontal beam is printed left to right (left column) the nozzle force causes a large deflection, resulting in elastic recoil once the force is removed. When the horizontal beam is printed right to left (right column), the deflection due to the nozzle force is small and the horizontal beam successfully fuses at both joints. All scale bars are 5 mm.}{
     \includegraphics[scale = .33]{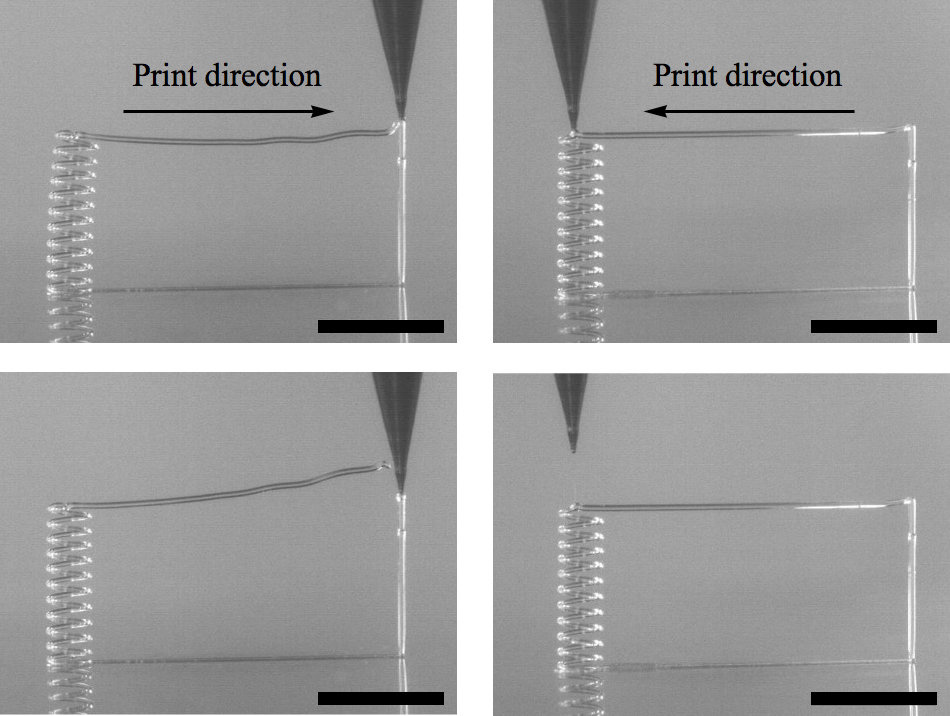}
   }
   
   We seek a printing plan that minimizes the probability that the print will fail. The probability that each beam incident to a given joint $q$ will successfully fuse is dependent on $\varepsilon_q$ as well as the material properties, the orientation of existing beams incident to that joint, and the printer operating parameters. We will assume simply that, if $\varepsilon_q$ is less than some critical distance $\varepsilon_c$, subsequent beams incident to $q$ will have $100\%$ probability of successfully fusing to $q$, and if $\varepsilon_q$ is greater than $\varepsilon_c$, then subsequent beams will have $0\%$ probability of successfully fusing to $q$. In this case there may be multiple optimal sequences in which $\varepsilon_q < \varepsilon_c$ for all $q$. Minimizing the maximum value of $\varepsilon_q$ will give one such sequence, if any exists. If in this solution the maximum value of $\varepsilon_q$ exceeds $\varepsilon_c$, then no solution exists, and no sequence will result in a successful print. For the purpose of computing this error, the free endpoint of this beam is coincident with the ending joint $q$ but not physically attached to any other beams incident \mbox{at $q$.}
   
   In order to estimate $\varepsilon_q$, we model the assembly as a linear, elastic frame.\cite{17} A vector of joint rotations and deflections $[\delta_i]$ is represented as the matrix product of a vector of applied forces and moments $[\delta_i]$ and a stiffness matrix $[k_{i,j}]$.\cite{33} This system of linear equations can be efficiently solved to find the resulting deflection at each joint in the assembly. For the designs tested, all the beams are the same diameter, and the beam diameter exceeds the nozzle diameter. Therefore we assume a constant, nominal load directed straight down, perpendicular to the nozzle orifice. We also assume that beam axes are straight lines, and approximate curved beams as sequences of straight beams. Finally, the weight of the beams themselves is neglected. The error $\varepsilon_q$ is simply the norm of the $x$, $y$, and $z$ deflections for the end joint \mbox{of the beam to be printed.}
   
   \myfigure{5}{t}{Illustration of the deflection computed using the exact (full frame) and heuristic (cantilever only) approaches.}{
     \includegraphics[scale = .18]{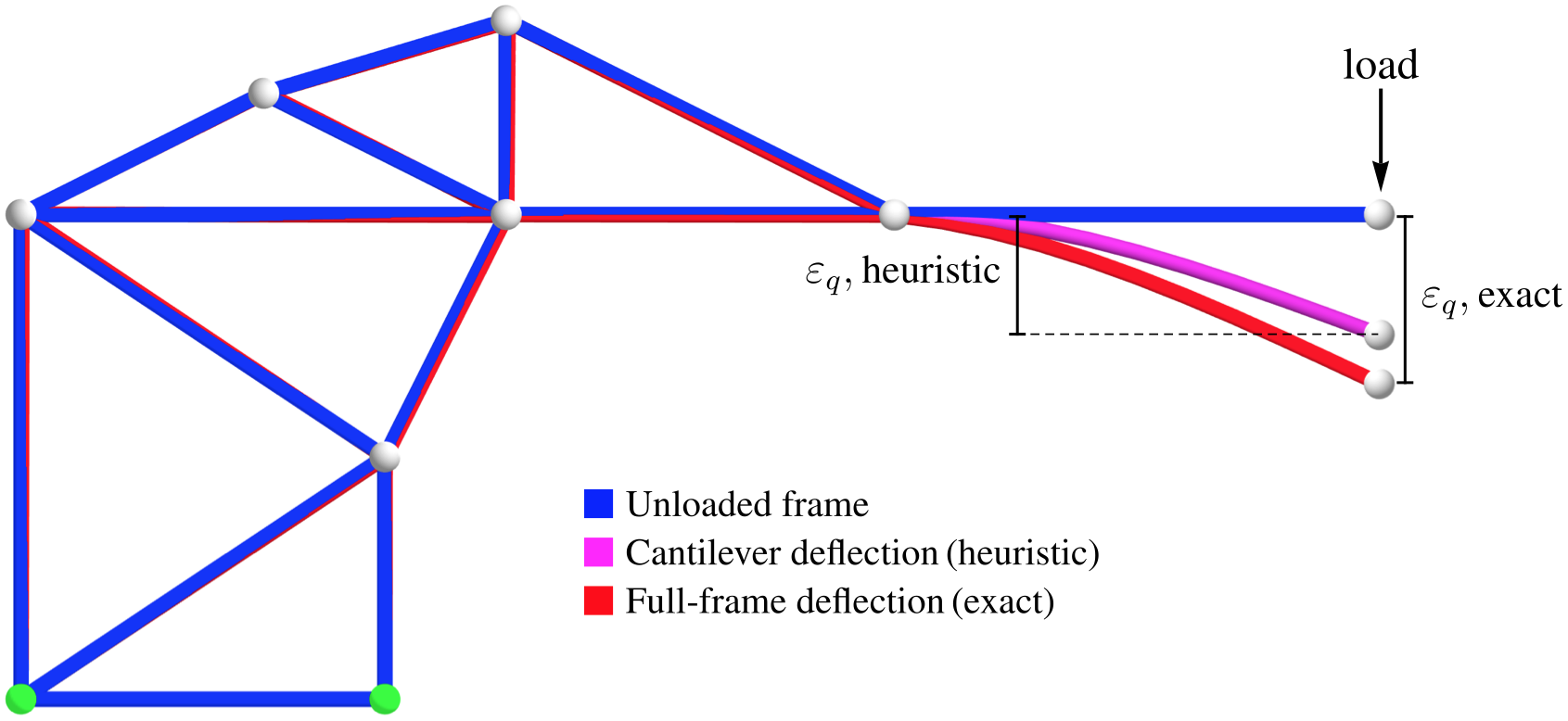}
   }
   
   Although direct stiffness methods are fast, modeling the deflection of the entire frame for every state explored can take substantial time for large assemblies. Therefore we also tested the following heuristic for $\varepsilon_q$ in our algorithm. When applying a load to the free end of a cantilevered beam in a freeform assembly, most of the deflection is usually due to bending and twisting of the cantilever. The non-cantilevered joints are incident to at least 2 beams instead of just 1 beam, so the rest of the frame is much stiffer, and the deflection and rotation of these non-cantilevered beams contributes minimally to the deflection at the end joint of the cantilever. Thus a low-cost state will usually be correctly identified by considering only the deflection of the cantilevered beam path and approximating the rest of the beams as being infinitely stiff. An example is shown in \cref{fig5}. This is a much faster computation than solving for the deflection of the whole frame, as 1) there are far fewer beams in the cantilevered set than in the entire frame and 2) the stiffness matrix is block diagonal, facilitating computation of its inverse.
   \vspace{27pt}
   
   \section{Planning Algorithm}
   \label{algorithm}
   
   In order to determine a sequence that minimizes the maximum cost of printing any given beam, we must first consider the relationship between the cost of a printing operation and the state in which it is performed. For any beam printed from joint $p$ to joint $q$, the addition of more beams to any part of the frame can only decrease $\varepsilon_q$ for that beam. If these additional beams are cantilevered themselves, they will have no effect. If these beams connect 2 existing joints in the frame, then they may have no effect or they may reduce $\varepsilon_q$. The cost associated with printing a new beam in a given direction in a given state is therefore an admissible heuristic for the cost of performing that operation in a later state. Less formally, the maximum cost of printing any beam can only be decreased by printing other beams before it.
   
Because every beam must be printed, and each beam can only decrease the maximum possible cost of all the remaining beams, the optimal sequence can be found by a best-first search. However, simply adding the lowest cost single beam at each step is not guaranteed to yield feasible sequence, because each joint may support only 1 cantilever in any given state. In order to guarantee that the cantilever constraint is not violated, we must instead add the lowest cost, constructable set of beams that 1) connects 2 or more existing, stable joints or 2) is cantilevered in the final assembly state. We define such a set of beams as follows.
   
   \begin{definition}
      Given current assembly state $G$ and goal state $H$, a \emph{consistent} subassembly $C(G,H)$ is a subassembly for which the following hold.
      \begin{enumerate}[label=\bfseries\arabic*.]
      \setlength{\parskip}{3pt}
        \item $G \cap C = \varnothing$.
        \item $Reachable(G, G \cup C)$ is true.
        \item No beam in $G \cup C$ is cantilevered, unless it is in $H$.
      \end{enumerate}
   \end{definition}
   
   The third condition guarantees that if $Reachable(G,H)$ is true, then $Reachable (G \cup C,H)$ is true. Given an algorithm for enumerating a consistent subassembly $C$ and a feasible sequence $C_{seq}$ connecting $G$ to $G \cup C$, a full assembly sequence for $H$ can be generated as in \cref{sequence2}.
   
   \begin{algorithm}
     \SetArgSty{}
     \DontPrintSemicolon
     \SetAlCapFnt{\footnotesize}
     \SetAlCapNameFnt{\footnotesize}
     \caption{A routine that finds a full assembly sequence for $H$ given an algorithm for finding a consistent subassembly $C$ and a feasible sequence $C_{seq}$. The sequence is recorded in $seq$.}
     \label{sequence2}
     $G \leftarrow \text{empty assembly state}$\;
     $H \leftarrow \text{full assembly state}$\;
     $seq \leftarrow \{\}$\;
     \While{$G \neq H$}{
       \If{there exists a consistent subassembly $C(G,H)$}{
         append $C_{seq}$ to $seq$\;
         $G \leftarrow G \cup C$\;
       }\Else{
         \Return unconstructable
       }
     }
     \Return $seq$\;
   \end{algorithm}
   
   Any algorithm that could enumerate a consistent subassembly sequence could be used to determine if an assembly were constructable, and hence could be used to decide circuit-SAT. Thus, any algorithm based on finding consistent subassemblies will in the worst case be intractable, unless P=NP. However, for most practical designs, consistent subassemblies can be identified by searching a relatively small space. An algorithm for enumerating the lowest-cost consistent subassembly sequence follows.
   
For each stable joint in the assembly, create a new subassembly state graph. In \cref{fig6}, the graphs are initialized with states $G_1$ and $G_2$ containing only the grounded green joints. Each subassembly will be cantilevered until it meets a different subassembly; thus, in order to respect the cantilever constraint, every state must consist of a simple beam path, and each subassembly state graph is a tree. At each step, add the beam $A$ to any state $G$ of any subassembly tree $T$ such that the following criteria are met.
   
   \begin{enumerate}[label=\bfseries\arabic*.]
     \setlength{\parskip}{2pt}
     \item Beam $A$ can be printed in state $G$ without violating a connection or cantilever constraint.
     \item All predecessors of $A$ are present either in $G$ or in any state of any different subassembly tree.
     \item The deflection caused by applying the load caused by the material flow at the endpoint of beam $A$ is the minimum of all feasible choices of $A$ not yet present in $T$.
   \end{enumerate}
   
   For each subassembly tree, track the joints that have been reached by any state in the tree. Eventually, 2 or more trees may reach a state containing a common joint, forming a path $S_1$ between their stable root joints. At this point, determine the set of all of the beams not in $S_1$ that must precede any of the beams in $S_1$. If this set is empty, $S_1$ is a consistent subassembly. In \cref{fig6}, this occurs when state $G_9$ is reached. $G_6 \cup G_9$ is a consistent subassembly.
   
   \myfigureexpl{6}{t}{Assembly algorithm}
   {In the first step (top), assembly trees $T_1$ and $T_2$ are generated starting from the green, stable joints in $G_1$ and $G_2$. The trees are expanded until there is a set of states $C$ such that $C$ contains at most 1 state from each tree such that the union of these states is a complete subassembly. In the first step, $C = G_6 \cup G_9$.\\ In the second step (bottom), the trees are reinitialized with the green, stable joints in states $G_{10}$ and $G_{11}$ and the process is repeated.}{
     \includegraphics[scale = .3]{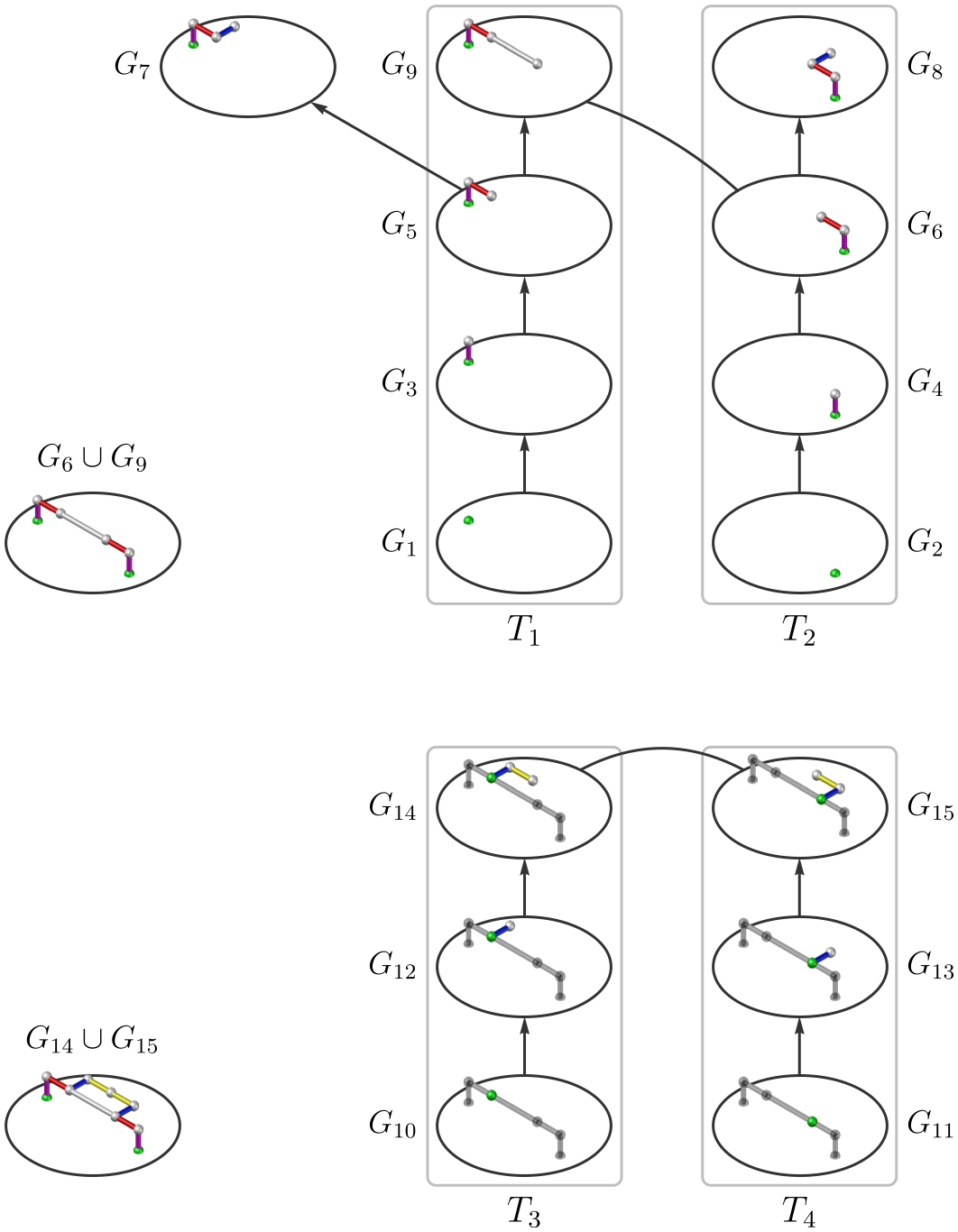}
   }
   
   If 2 or more states in different subassembly trees reach a common joint, but there are still predecessors of $S_1$ that are not in $S_1$, the search must continue until there is a set of paths $C$ such that all of the predecessors of all beams in $C$ are contained in $C$. If this condition is satisfied, then $Reachable(G, G \cup C)$ is true, and $C$ is a consistent subassembly. Since $C$ may contain any number of paths, this step has worst-case exponential run time. For example, for an assembly that instantiates circuit-SAT as in the proof above, any consistent subassembly requires assembly of the output beam, which requires assembly of exactly 1 of the red or blue beam paths for each gadget. Thus, for $n$ gadgets, the worst-case run time can be no better than $\Omega(2^n)$.
   
   Once the consistent subassembly $C$ is found, record the partial ordering constraint that every beam in $C$ must follow every beam present in the connected component to which $C$ is added. Also record a directionality constraint such that each beam must be printed in the direction corresponding to the order in which the search visited its joints. This ensures that the 2 cantilevered sets that are joined to make each path $S$ always meet at the same joint $q$. As long as the 2 sides of $S$ meet at $q$, the maximum error for any joint in $S$ will be the same regardless of the actual sequence in which $S$ is printed.
   
After adding a consistent subassembly, introduce new subassembly trees for the new stable joints. In \cref{fig6}, these trees are rooted at states $G_{10}$ and $G_{11}$. Delete any states containing only beams that are present in the new consistent subassembly. In \cref{fig6}, this will remove states $G_1$ through $G_6$ and $G_9$. For each remaining state $G$ containing one of the new stable joints $p$, replace $G$ with $G \setminus C$ and reassign it as a child of joint $p$. In \cref{fig6}, for example, $G_{12}$ is created from $G_7 \setminus (G_6 \cup G_9)$ and is rooted at the green joint in $G_{10}$, and $G_{13}$ is created from $G_8 \setminus (G_6 \cup G_9)$ and is rooted at the green joint in $G_{11}$.
   
In \cref{fig6}, the cost associated with state transition \mbox{$G_{10} \rightarrow G_{12}$} is less than the cost associated with $G_5 \rightarrow G_7$, even though each transition corresponds to the same beam being assembled. In general, because the new consistent subassembly can change the stiffness matrix representing its connected subassembly, the cost for each state transition in this connected subassembly must be recomputed. The cost of state transitions in different connected subassemblies remains the same and need not be recomputed.
   
This process is repeated until the all the beams are part of a consistent subassembly. At any point, if all the predecessors of a beam $A$ that is cantilevered in the full assembly state have been assembled, $A$ can be assembled. No additional partial ordering is associated with $A$. Also, because the deflection of existing joints does not appear in the cost function, if any 2 existing joints can be joined by a single beam, the cost of the consistent subassembly corresponding to \mbox{that single beam is 0.} In these cases, a consistent subassembly will consist of a single beam with 0 cost.
   
Any print sequence that is consistent with all of the precedence constraints enumerated by the previous steps will result in the same maximum cost. More importantly, a search of the assembly graph subject to these additional constraints is now guaranteed not to reach a dead end. Some further optimization can be performed within these constraints with respect to the time spent on non-printing motion, termed \emph{deadheading}. As before, we will search for a path through assembly states. However, for this assembly graph search, the assembly state is specified by the set of beams that has been printed as well as the position of the nozzle. Thus there are two possible state transitions: a printing transition and a deadhead transition. The cost of a deadheading state transition is the time required to move the nozzle from the end joint of one beam to the start joint of the next beam in the sequence. Because every printing step will take the same amount of time, regardless of the sequence, the cost associated with a printing state transition is irrelevant to the time optimization problem and may be assigned a value of 0. It is therefore possible to apply a best-first search to find a time-optimal assembly sequence with respect to deadheading time alone. Determining the minimum possible cost of any single deadheading transition requires computing the fastest trajectory between the start and end joints that avoids collision with any of the beams that have already been printed. This is a classical motion planning problem, and determination of a global optimum would require solving this problem for each possible deadheading state transition at each step. However, because deadheading moves can be performed at speeds orders of magnitude faster than printing speeds, such a rigorous optimization would provide only marginal time savings over a simpler approach. We avoided deadheading collisions simply by raising the nozzle above the maximal $z$ value of any beam in the assembly state, moving it to the $xy$ coordinates of the next joint, then lowering it to the $z$ value of the next joint. Given the maximum speed of each axis, and assuming instantaneous acceleration, the deadheading time is easily computed. A greedy search of the assembly graph using this time as the cost function, subject to the constraints enumerated during the search for complete subassemblies, provided satisfactory assembly sequences for the designs shown here.
   
   \section{Algorithm Performance and Physical Validation}
   \label{performance}
   
   To validate our model and algorithm, we printed three complex geometries: the ArcelorMittal Orbit (2,201 beams), a mesh approximation of the human heart (4,637 beams), and a moderately simplified version of the Eiffel tower (6,161 beams). The Orbit took 5 hours to print, the heart took 8 hours to print, and the Eiffel tower took 13 hours to print. A printing video of a smaller design, a Pennsylvania bridge, is shown as well in the supplement (supplementary video 2). \autoref{fig7} shows the successfully printed designs. \autoref{fig8} shows the performance of the assembly sequencing algorithm using the two different methods of computing the cost function.
   
   \myfigure{7}{t}{3D printed designs. All scale bars are 5 mm.}{
     \includegraphics[scale = .95]{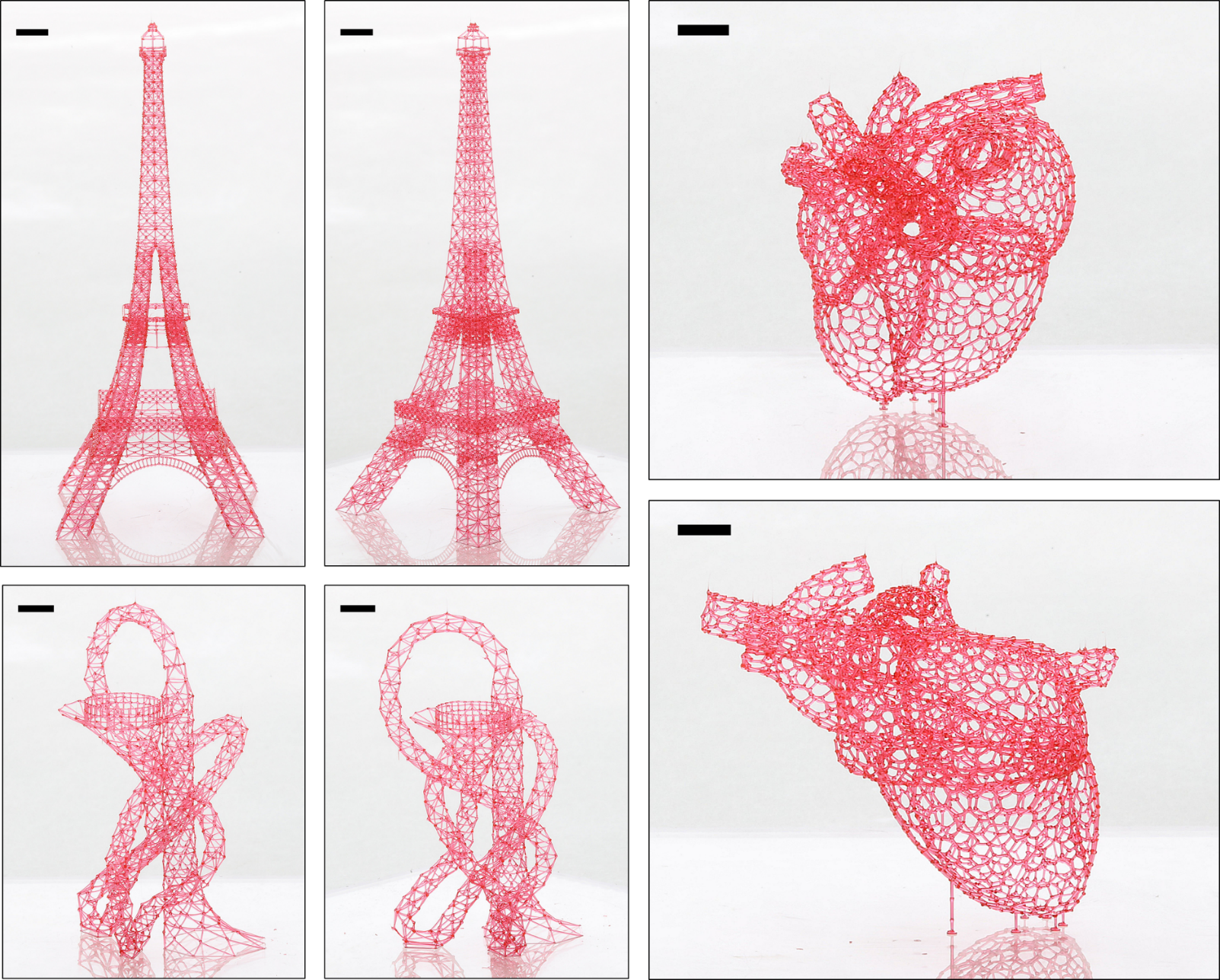}
   }
   
   Assembly sequences were generated using both the heuristic and exact cost functions for all three designs. The exact cost was then computed for each state in the ``heuristic'' assembly sequence. \autoref{fig8}A shows the exact compliance (error normalized by applied load) for all states in the sequences generated using each algorithm, sorted from smallest to largest. The exact compliances encountered in the heuristic sequence are slightly greater than the exact compliances found in the exact sequence, indicating that the heuristic algorithm finds a slightly worse solution than the exact one. \autoref{fig8}B shows the time required to compute each sequence for the three designs. \autoref{fig8}C shows the correlation between the heuristic and exact compliances for all the states in the heuristic sequence. As expected, the compliance estimated using the heuristic is always less than or equal to the exact compliance. Both sequences were successfully used to print all three designs.
   
   \myfigureexpl{8}{t}{Algorithm performance}{A - Distribution of compliances in sequences generated using the heuristic and exact cost functions. For each state in the print sequence generated using the heuristic cost function, the exact cost function is computed as well. The exact cost functions are then normalized by the applied load to give compliances. Solid lines correspond to the sequence computed using the exact cost function; dashed lines correspond to the sequence generated using the heuristic cost function.\\
   B - Sequencing times using both heuristic and exact cost functions for all three designs.\\
   C - Correlation between heuristic and exact compliances for all states in the assembly sequence generated using the heuristic cost function.}{
     \includegraphics[scale = .2925]{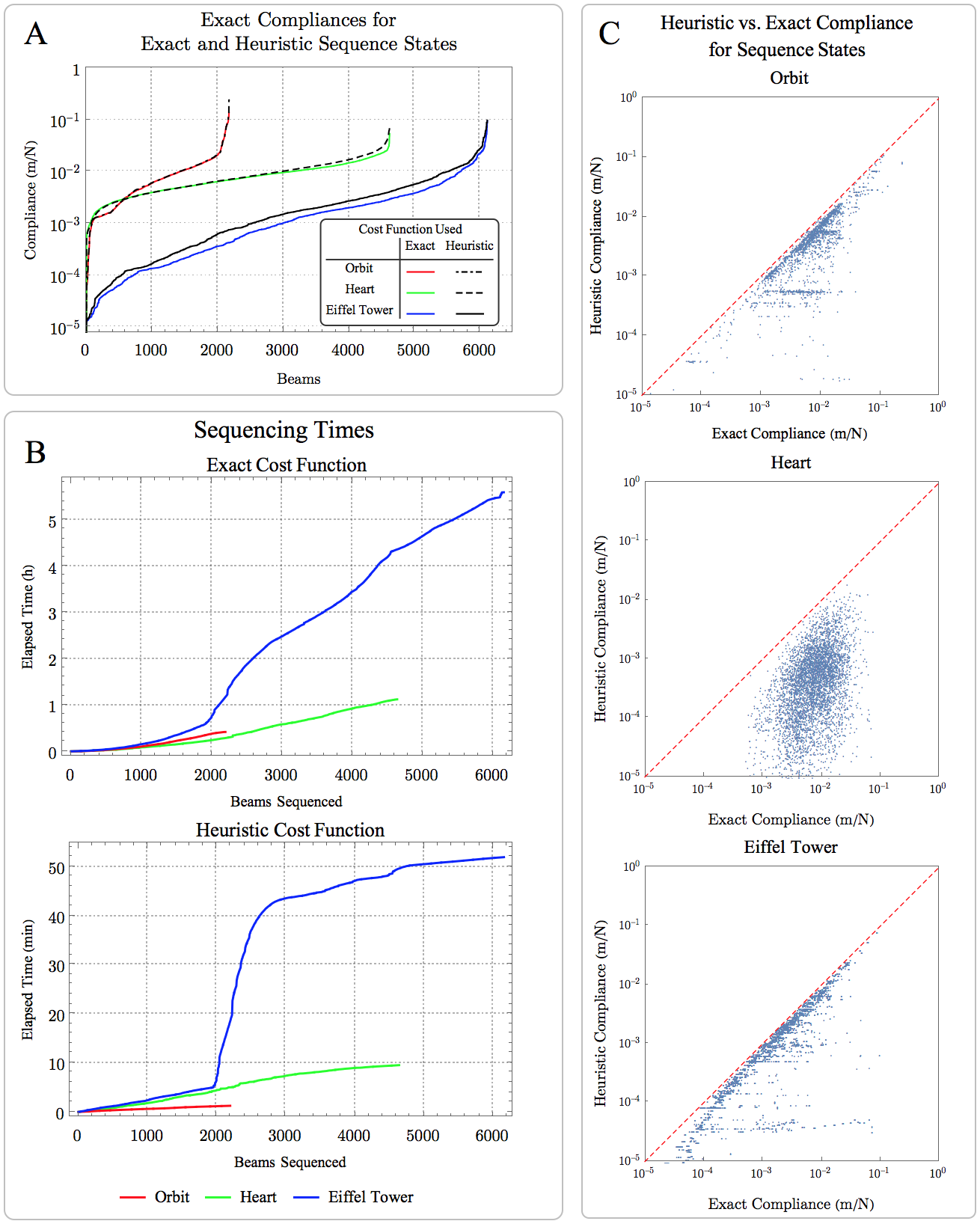}
   }
   
The timing for all three constructs was dominated by the time required to compute the deflections. While obtaining a truly optimal sequence requires solving this system of equations for every joint at every step, near-optimal sequences can be obtained by determining the deflection assuming that each root joint of each subassembly tree is fixed, excluding all non-cantilevered beams from the computation. This allows generation of assembly sequences in a much shorter time than that required to physically print the design. However, even using the exact cost function, planning required less time than the actual printing process for these designs. Because sequencing is initiated from the empty state, the planner could be run on the fly without the printer ever waiting on the planner. For even larger designs, however, sequencing using the exact cost function would take longer than printing, and it would be advantageous to use the heuristic cost function instead.
   
   \section{Discussion}
   \label{discussion}
   
   We have developed and validated an approach to generating mechanically robust assembly sequences for freeform assembly. The algorithm generates an optimal assembly plan for specific cost functions, but it can also be used to generate good assembly plans under more general cost functions. Even if an assembly process permits multiple cantilevers at the same joint, for any cost function related to deformation or internal beam forces, it is generally preferable to minimize the number and length of cantilevers present in any given state. The algorithm does this by adding the lowest-cost consistent subassembly, rather than the lowest-cost single beam. If the applied load is due to the self-weight of the assembly, adding the lowest-cost single beam will yield a suboptimal assembly sequence, as shown in an example structure given by McEvoy \etal\cite{27} Though the algorithm presented here will not in all cases return a global optimum for loading conditions due to the beams' self-weight, it returns the optimal sequence for that particular example. Thus, this approach may be useful for robotic assembly of frame structures from pre-formed beams as well as for variants of freeform assembly to which the cantilever constraint does not strictly apply.

We have addressed this assembly problem with the cantilever constraint using a 3-axis printer for a design in a set orientation. However, designs that are unconstructable in a given orientation on a 3-axis printer may be constructable in a different orientation. A logical next step in this work is to develop an algorithm that finds this orientation. There are also designs which are unconstructable in any orientation, in which case it becomes necessary to consider a printer with more degrees of freedom.

Recent work has explored assembly planning for a wireframe assembly process under collision and connection constraints for robots with 5 (Wu\cite{16}) and 6 (Huang\cite{17}) degrees of freedom. Wu \etal implement a planner that avoids collision and unnecessary deadheading without considering stability. They use a ``peeling'' method that cuts the design into a series of subassemblies, for which a collision-free ordering can always be found if it exists. This could be considered a divide-and-conquer approach. However, it is possible to for the algorithm to generate subassemblies that cannot be completed without collision, even though there exists a feasible ordering for the entire assembly. This would easily be resolved using a disassembly approach as described in our introduction. With a disassembly state space search, under collision and connection constraints only, it is impossible to reach a dead end. If deciding whether a single beam can be assembled in a given state can be done in polynomial time, then so can generation of the full sequence.

Huang \etal address a similar problem, but they introduce a stability criterion involving the maximum deflection of each joint under gravitational force. Though not proven, because this criterion is a function of the assembly state, it seems very likely that deciding the existence of a feasible sequence is NP-complete. The authors also perform some heuristic optimization for a cost function that is a weighted sum of the maximum deflection of a joint due to the frame's self-weight, deadheading, and the probability that a beam will prevent the placement of another beam later in the assembly due to the collision constraint. In contrast, our cost function is the deflection of a joint due to viscous coupling between the nozzle and the extruded material; this is modeled as a point force at the nozzle, which results in very different and much larger deflections than those due to the assembly's self-weight. Additionally, our process is subject to the cantilever constraint. Due to the scale of the beams and the thermal properties of isomalt, placing the nozzle at a joint inevitably melts the material there. This is not apparently not true of the ABS carbon powder composite used by Huang and the unspecified thermoplastic used by Wu. While these implementations tolerate multiple cantilevers, our process does not. This constraint renders the problem of finding a feasible sequence NP-complete and precludes the application of existing algorithms.

The planners used by Wu and Huang used a forward assembly search, but in both cases, a disassembly search would likely enable a simpler approach. Once a beam is disassembled in a disassembly graph search, it is no longer necessary to compute collision constraints for that beam. For a 3-axis system, the precedence relationships between beams are the same regardless of the assembly state. It was therefore feasible for us to compute all of them before starting the algorithm. If we had implemented a disassembly approach and computed collision constraints as needed, we would only need to compute half as many constraints as we did for the forward search. The benefits of a forward search (easier conceptualization, concurrent planning and printing) outweighed the computational cost of computing these additional constraints, which were trivial compared to other steps in the algorithm. However, for a printer with 5 or 6 axes, the precedence relationship between any 2 beams depends on the assembly state. It cannot be known, as it can for a 3-axis printer, whether assembly of beam $A$ will obstruct the assembly of a beam $B$. Disassembly of a beam $A$, in contrast, can only make it easier to disassemble beam $B$. Equivalently, if a beam can be disassembled in state $H$, then it can surely be disassembled in state $G \subset H$. The algorithms used by both Wu and Huang rely on a heuristic for cutting the assemblies into subassemblies that can be sequenced by a forward planner. In both cases, it is possible to generate subassemblies that are unconstructable due to collision constraints, even though a feasible sequence for the entire assembly exists. This is equivalent to reaching a dead end in a forward search, and would be avoided entirely by a disassembly search. The problem might also then be more amenable to existing best-first search algorithms such as A*, which have been successfully applied to disassembly problems\cite{34}.
   
Our primary motivation for exploring freeform assembly was the applications of round-channel microchannel networks that can be formed by using the freeform constructs as sacrificial molds. In the sacrificial molding process, a freeform 3D printed template is cast in a curable material. Dissolving or otherwise removing the template yields a 3D network of cylindrical channels. This can be used to make, for example, 3D microfluidic devices in materials with tunable optical properties\cite{35}. A similar approach could be used to create woven or interpenetrating channel networks in structural polymers, enabling efficient delivery of 2-part healing chemistries to cracks.\cite{36,37} The ability to incorporate 3D microchannel networks in elastomeric polymers also suggests applications in integrated microfluidics and soft robotics.\cite{38} Round-channel networks are also useful in recapitulating the structure of native tissue.\cite{39,40,41,42,43,44} Breast and prostate, for example, contain independent networks of lymphatic, vascular, and ductal channels. Carcinomas develop in the ducts and metastasize \emph{via} either the lymphatic or vascular system; thus, a complete ``organ-on-a-chip'' model requires all 3 networks. Freeform assembly can be used to generate this complex topology in a variety of hydrogels, and the channels can be seeded in a subsequent step.\cite{45}
   
Extrusion based bioprinting can directly pattern cells in a single step. If cells are patterned into a supporting reservoir,\cite{4,5} the geometry is relatively unconstrained. However, it is yet unclear how such an approach would enable oxygen delivery to cells within the reservoir during long prints. If cells are patterned layer-by-layer, the construct can be made porous, so that oxygen demand can be met during printing by diffusion from the gas phase. Printing designs with pores or microchannels is facilitated by incorporation of a stiff polymer into the construct,\cite{46} but it is possible to print a porous construct using only cell-laden gels as well.\cite{47} In the latter case, the pores are backfilled, and some of the patterned gels are liquefied and removed to create a different vascular network, better suited to perfusion with liquid media. The chief limitation of these layer-by-layer approaches is that the gels have low stiffness and must be patterned without cantilevers or long unsupported spans. This limits the topology of the channel networks that can be created. A highly cited but relatively undeveloped alternative approach is to print an acellular sacrificial template using water-soluble carbohydrate glass\cite{19,48} or polymer,\cite{49} then cast a cell-laden hydrogel around it in a single, rapid step. Freeform assembly with a heated nozzle is a powerful means of patterning these stiff, biocompatible, water-soluble materials, but improved printing precision and a solution to the sequencing problem were required to extend this approach to arbitrarily large, complex designs.
   
The designs printed here were generated by essentially tracing existing designs or from meshing of non-uniform rational basis spline surfaces, with some manual tweaking. Thus, the fully assembled state was specified from the start of the program. However, a similar graph search approach could be used to generate the topology concurrently with the sequence. In this case the objective function for any candidate beam would have to reflect not only the cost (deflection) that would occur during assembly, but also the benefit of the candidate beam or subassembly with respect to the ultimate function of the design, i.e., transport of mass or heat.
   
   \section{Materials and Methods}
   \label{methods}
   
   \subsection{Design Generation}
   The heart design was based off a Solidworks model of a human heart uploaded to GrabCAD by M.G. Fouch\'e. The surfaces comprising the heart were trimmed and patched to leave a set of non-manifold surfaces. The SolidWorks Simulation package was used to generate a triangular mesh for each surface. The mesh was exported to Mathematica and the centroid of each triangle was connected to form the dual. The resulting graph was then exported to AutoCAD, where the connected components were manually joined and the supports were manually added.
   
The design of the ArcelorMittal Orbit was based off a Sketchup model uploaded to 3D Warehouse by user Damo. The file was converted to .dxf and modified in AutoCAD to generate a design suitable for printing. Some pairs of incident beams in the Orbit were both oriented within about 13\textdegree\,\,of the vertical, which would make collision unavoidable for the shape of the nozzles used on the printer. To make the design constructable, the vertical dimensions of the Orbit were scaled down to $85\%$ of their original value. The design of the Eiffel tower was based off a .dwg file uploaded to BiblioCAD by user limazkan. The geometry was used as a guide to create a printable design in AutoCAD.
   
   \subsection{Algorithm Implementation}
   The search algorithm and direct stiffness method were implemented in the Wolfram Language. The assembly was modeled as a linear, elastic frame with elements of circular cross-section. The elastic modulus was set to 2.6 GPa\cite{20} and the shear modulus was set to 1.1. This is based off the assumption that isomalt is linear and isotropic and has a Poisson's ratio of 0.2, a typical value for glasses. A nominal load of $10^{-4}$ Newtons was applied to determine the stiffness of each joint. The systems of linear equations were solved with the Wolfram Language command \href{http://reference.wolfram.com/language/ref/LinearSolve.html}{\texttt{LinearSolve}}. All sequencing was performed on a machine with a quad core 2.8 GHz Intel Core i7 processor and 16 GB of RAM.
   
   \subsection{Isomalt Processing}
   Isomalt has a tendency to co-crystallize with water under the heat and shear stress inherent in melt extrusion processes. This can be prevented by drying the isomalt prior to printing. 50 grams of isomalt (GalenIQ 990, Beneo-Palatinit Gmbh), 25 mg Allura Red AC (Sigma), and a 3\textquotedbl \,PTFE stir bar were placed in a 250 mL vacuum flask. The flask was immersed in a temperature controlled oil bath. The bath was raised to 100\textdegree C. Vacuum was applied using a rotary vane pump. The bath temperature was then ramped to 150\textdegree C at a rate of 120\textdegree C/hr. The isomalt was held at 150 \textdegree C and stirred at 120 RPM for 2 hours, then cast into an aluminum mold to make a solid rod. The rod was stored under nitrogen until printing.
   
   \subsection{Isomalt Printing}
   The isomalt was printed using a custom printer as described previously\cite{20,35}, using a 110\;$\mu$m nozzle (Subrex). About 10 g of isomalt was loaded into the extruder barrel and covered with a layer of pump oil. Designs were printed at 400\;$\mu$m/s at a pressure of 0.6-1.2 mPa. The printer has 2 temperature control zones corresponding to the nozzle and the extruder barrel. The nozzle zone was set to 150\textdegree C and the barrel zone was set to 105\textdegree C. The designs were printed onto a substrate of Polyethylene terephthalate, glycol-modified (PETG). The printer enclosure was continuously purged with dry air from the central compressed air line, maintaining a relative humidity under $20\%$. After printing, constructs were stored under nitrogen. However, constructs were photographed under ambient conditions and showed no apparent degradation over the course of several hours.
   
   \section*{Acknowledgments}
   \label{acknowledgments}
   
   M. K. Gelber was supported by fellowships from the Roy G. Carver Foundation and the Arnold and Mabel Beckman Foundation. We gratefully acknowledge the gift of isomalt and advice on its processing provided by Oliver Luhn of S\"udzucker AG/Beneo-Palatinit GmbH. The development of the printer was supported by the Beckman Institute for Advanced Science and Technology \emph{via} its seed grant program.
   
We also would like to acknowledge Travis Ross of the Beckman Institute Visualization laboratory for help with macro photography of the printed constructs. We also thank the contributors of the CAD files on which we based our designs: GrabCAD user M.G. Fouch\'e, 3DWarehouse user Damo, and BiblioCAD user limazkan (Javier Mdz). Finally, we acknowledge Seth Kenkel for valuable feedback on the manuscript.


\begin{thebibliography}{99}
     
   \bibitem{1}
     Calignano, F. \etal Overview on Additive Manufacturing Technologies. 
     \emph{Proceedings of the IEEE} {\bf 105}, 593\dash612, \doi{10.1109/jproc.2016.2625098} (2017).

   \bibitem{2}
     Wu, W. \etal. Direct-write assembly of biomimetic microvascular networks for efficient fluid transport. 
     \emph{Soft Matter} {\bf 6}, 739\dash742, \doi{10.1039/b918436h} (2010).
     
   \bibitem{3}
     Wu, W., DeConinck, A. \& Lewis, J. A. Omnidirectional Printing of 3D Microvascular Networks. 
     \emph{Adv. Mater.} {\bf 23}, H178\dash{H183}, \doi{10.1002/adma.201004625} (2011).
     
   \bibitem{4}
     Bhattacharjee, T. \etal Writing in the granular gel medium. 
     \emph{Science Advances} {\bf 1},\\ \doi{10.1126/sciadv.1500655} (2015).
     
   \bibitem{5}
     Hinton, T. J. \etal Three-dimensional printing of complex biological structures by freeform reversible embedding of suspended hydrogels. 
     \emph{Science Advances} {\bf 1}, \doi{10.1126/sciadv.1500758} (2015).
   
   \bibitem{6}
      O'Bryan, C. S. \etal Self-assembled micro-organogels for 3D printing silicone structures. 
      \emph{Science Advances} {\bf 3}, \doi{10.1126/sciadv.1602800} (2017).
   
   \bibitem{7}
      Lebel, L. L., Aissa, B., El Khakani, M. A. \& Therriault, D. Ultraviolet-assisted direct-write fabrication of carbon nanotube/polymer nanocomposite microcoils. 
      \emph{Adv. Mater.} {\bf 22}, 592\dash596, \doi{10.1002/adma.200902192} (2010).
     
   \bibitem{8}
     Lu, Y., Vatani, M. \& Choi, J.-W. Direct-write/cure conductive polymer nanocomposites for 3D structural electronics. 
     \emph{J Mech Sci Technol} {\bf 27}, 2929\dash2934, \doi{10.1007/s12206-013-0805-4} (2013).
   
   \bibitem{9}
     \emph{A Radically New 3D Printing Method} \url{http://www.mataerial.com} (December 2017)
     
   \bibitem{10}
     Guo, S. Z. \etal Solvent-cast three-dimensional printing of multifunctional microsystems. 
     \emph{Small} {\bf 9}, 4118\dash4122, \doi{10.1002/smll.201300975} (2013).
   
   \bibitem{11}
     Ahn, B. Y. \etal Omnidirectional Printing of Flexible, Stretchable, and Spanning Silver Microelectrodes. 
     \emph{Science} {\bf 323}, 1590\dash1593, \doi{10.1126/science.1168375} (2009).
   
   \bibitem{12}
     Skylar-Scott, M. A., Gunasekaran, S. \& Lewis, J. A. Laser-assisted direct ink writing of planar and 3D metal architectures. 
     \emph{PNAS} (2016).
   
   \bibitem{13}
     Hack, N. \& Lauer, W. V. Mesh-Mould: Robotically Fabricated Spatial Meshes as Reinforced Concrete Formwork. 
     \emph{Architectural Design} {\bf 84}, 44\dash53, \doi{10.1002/ad.1753} (2014).
   
   \bibitem{14}
     Mueller, S. \etal in \emph{Proceedings of the 27th annual ACM symposium on User interface software and technology} 273\dash280 (ACM, Honolulu, Hawaii, USA, 2014).
   
   \bibitem{15}
     Peng, Huaishu \etal On-The-Fly Print: Incremental Printing While Modelling. \emph{Proceedings of the 2016 CHI Conference on Human Factors in Computing Systems}, 887\dash896, \doi{10.1145/2858036.2858106} (2016).
     
   \bibitem{16}
     Wu, Rundong \etal Printing Arbitrary Meshes with a 5DOF Wireframe Printer. \emph{ACM Trans. Graph} {\bf 35}, 4, 101:1\dash101:9, \doi{10.1145/2897824.2925966} (2016).
     
   \bibitem{17}
     Huang, Jiyang \etal FrameFab: Robotic Fabrication of Frame Shapes. \emph{ACM Trans. Graph} {\bf 35}, 6, 224:1\dash224:11, \doi{10.1145/2980179.2982401} (2016).
     
   \bibitem{18}
     \emph{Printing outside the box}, \url{http://mx3d.com/} (December 2017)
   
   \bibitem{19}
     Miller, J. S. \etal Rapid casting of patterned vascular networks for perfusable engineered
three-dimensional tissues. \emph{Nat Mater} {\bf 11}, 768\dash774 (2012).
   
   \bibitem{20}
     Gelber, M. K. \& Bhargava, R. Monolithic multilayer microfluidics via sacrificial molding
of 3D-printed isomalt. \emph{Lab. Chip} {\bf 15}, 1736\dash1741, \doi{10.1039/c4lc01392a} (2015).
     
   \bibitem{21}
     Gelber, M. K. \etal Quantitative Chemical Imaging of Nonplanar Microfluidics. \emph{Analytical Chemistry} {\bf 89}, 3, 1716\dash1723, \doi{10.1021/acs.analchem.6b03943} (2017).
   
   \bibitem{22}
     Romney, B., Godard, C., Goldwasser, M. \& Ramkumar, G. in \emph{ASME Database
Symposium}. 699\dash712.
     
   \bibitem{23}
     Homem de Mello, L. \& Sanderson, A. C. A correct and complete algorithm for the generation of mechanical assembly sequences. \emph{Robotics and Automation, IEEE Transactions on} {\bf 7}, 228\dash240 (1991).

   \bibitem{24}
     Homem de Mello, L. S. \& Sanderson, A. C. AND/OR graph representation of assembly plans. 
     \emph{Robotics and Automation, IEEE Transactions on} {\bf 6}, 188\dash199, \doi{10.1109/70.54734} (1990).
     
   \bibitem{25}
     Elliott Fahlman, S. A planning system for robot construction tasks. 
     \emph{Artificial Intelligence} {\bf 5}, 1\dash49, \doi{10.1016/0004-3702(74)90008-3} (1974).
     
   \bibitem{26}
     Schmult, B. Autonomous Robotic Disassembly in the Blocks World. 
     \emph{The International Journal of Robotics Research} {\bf 11}, 437\dash459, \doi{10.1177/027836499201100502} (1992).
     
   \bibitem{27}
     McEvoy, M., Komendera, E. \& Correll, N. in \emph{Technologies for Practical Robot Applications (TePRA), 2014 IEEE International Conference on}. 1\dash6.
   
   \bibitem{28}
      Homem de Mello, L. S. \& Desai, R. S. Assembly planning for large truss structures in space. 
      \emph{Systems Engineering, 1990., IEEE International Conference on}, 404\dash407, \doi{10.1109/icsyse.1990.203182} (1990).
   
   \bibitem{29}
      Abe, S., Murayama, T., Oba, F. \& Narutaki, N. in \emph{Systems, Man, and Cybernetics, 1999. IEEE SMC '99 Conference Proceedings. 1999 IEEE International Conference on}. 486\dash491 vol. 482.
     
   \bibitem{30}
     Mosemann, H., R\"ohrdanz, F. \& Wahl, F. in \emph{Robotics and Automation, 1998. Proceedings. 1998 IEEE International Conference on}. 233\dash238 (IEEE).
     
   \bibitem{31}
     M\"ohring, R. H., Skutella, M. \& Stork, F. Scheduling with AND/OR precedence constraints. 
     \emph{SIAM Journal on Computing} {\bf 33}, 393\dash415 (2004).
   
   \bibitem{32}
     Goldschlager, L. M. The monotone and planar circuit value problems are log space complete for P. 
     \emph{SIGACT News} {\bf 9}, 25\dash29, \doi{10.1145/1008354.1008356} (1977).
   
   \bibitem{33}
     Budynas, R. G. \emph{Advanced Strength and Applied Stress Analysis}. (McGraw-Hill, 1999).
   
   \bibitem{34}
     Laperri\`ere, L. and ElMaraghy, H.A. Planning of Products Assembly and Disassembly. \emph{CIRP Annals} {\bf 41}, 1, 5\dash9, \doi{10.1016/S0007-8506(07)61141-X} (1992).

   \bibitem{35}
     Gelber, M. K., Kole, M. R., Kim, N., Aluru, N. R. \& Bhargava, R. Quantitative Chemical Imaging of Non-Planar Microfluidics. 
     \emph{Anal Chem}, \doi{10.1021/acs.analchem.6b03943} (2016).
   
   \bibitem{36}
     Hansen, C. J. et al. Self-Healing Materials with Interpenetrating Microvascular Networks. 
     \emph{Adv. Mater}. {\bf 21}, 4143\dash4147, \doi{10.1002/adma.200900588} (2009).
   
   \bibitem{37}
     Hamilton, A. R., Sottos, N. R. \& White, S. R. Self-Healing of Internal Damage in Synthetic Vascular Materials. 
     \emph{Adv. Mater}. {\bf 22}, 5159\dash5163, \doi{10.1002/adma.201002561} (2010).
   
   \bibitem{38}
     Wehner, M. \etal An integrated design and fabrication strategy for entirely soft, autonomous robots. 
     \emph{Nature} {\bf 536}, 451\dash466 (2016).
   
   \bibitem{39}
     Huang, Z., Li, X., Martins-Green, M. \& Liu, Y. Microfabrication of cylindrical microfluidic channel networks for microvascular research. \emph{Biomed. Microdevices} {\bf 14}, 873\dash883, \doi{10.1007/s10544-012-9667-2} (2012).
   
   \bibitem{40}
     Yang, X., Forouzan, O., Burns, J. M. \& Shevkoplyas, S. S. Traffic of leukocytes in microfluidic channels with rectangular and rounded cross-sections. \emph{Lab. Chip} {\bf 11}, 3231\dash3240, \doi{10.1039/c1lc20293f} (2011).
     
   \bibitem{41}
     Fiddes, L. K. \etal A circular cross-section PDMS microfluidics system for replication of cardiovascular flow conditions. 
     \emph{Biomaterials} {\bf 31}, 3459\dash3464, \doi{10.1016/j.biomaterials.2010.01.082} (2010).

   \bibitem{42}
     Bischel, L. L., Young, E. W. K., Mader, B. R. \& Beebe, D. J. Tubeless microfluidic angiogenesis assay with three-dimensional endothelial-lined microvessels. \emph{Biomaterials} {\bf 34}, 1471\dash1477, \doi{10.1016/j.biomaterials.2012.11.005} (2013).
     
   \bibitem{43}
     Dolega, M. E. \etal Facile bench-top fabrication of enclosed circular microchannels provides 3D confined structure for growth of prostate epithelial cells. \emph{PLoS One} {\bf 9}, \doi{10.1371/journal.pone.0099416} (2014).
     
   \bibitem{44}
     Bischel, L. L., Beebe, D. J. \& Sung, K. E. Microfluidic model of ductal carcinoma in situ with 3D, organotypic structure. 
     \emph{BMC Cancer} {\bf 15}, 12 (2015).
     
   \bibitem{45}
     Kinstlinger, I. S. \& Miller, J. S. 3D-printed fluidic networks as vasculature for engineered tissue. 
     \emph{Lab. Chip} {\bf 16}, 2025\dash2043, \doi{10.1039/c6lc00193a} (2016).
   
   \bibitem{46}
      Kang, H.-W. \etal A 3D bioprinting system to produce human-scale tissue constructs with structural integrity. 
      \emph{Nat Biotech} {\bf 34}, 312\dash319, \doi{10.1038/nbt.3413} (2016).
   
   \bibitem{47}
      Kolesky, D. B., Homan, K. A., Skylar-Scott, M. A. \& Lewis, J. A. Three-dimensional bioprinting of thick vascularized tissues. 
      \emph{Proc. Natl. Acad. Sci. U.S.A}. {\bf 113}, 3179\dash3184, \doi{10.1073/pnas.1521342113} (2016).
     
   \bibitem{48}
     Bellan, L. M. \etal Fabrication of an artificial 3-dimensional vascular network using sacrificial sugar structures. 
     \emph{Soft Matter} {\bf 5}, 1354\dash1357, \doi{10.1039/b819905a} (2009).
     
   \bibitem{49}
     Mohanty, S. \etal Fabrication of scalable and structured tissue engineering scaffolds using water dissolvable sacrificial 3D printed moulds. \emph{Materials Science and Engineering: C} {\bf 55}, 569\dash578, \doi{10.1016/j.msec.2015.06.002} (2015).
     
   \end{thebibliography}
\end{document}